\documentclass{icrc}

\usepackage{times}
\usepackage{graphicx} 

\begin{document}

\title{Gamma-Hadron Separation using \v Cerenkov Photon Density Fluctuations}
\author[1]{P. N. Bhat}
\author[1]{V. R. Chitnis}
\affil[1]{Tata Institute of Fundamental Research, Homi Bhabha Road,
Mumbai 400 005, India}

\correspondence{P. N. Bhat (pnbhat@tifr.res.in)}

\runninghead{Bhat and Chitnis : Gamma-Hadron Separation using Density Fluctuations }
\firstpage{1}
\pubyear{2001}


\maketitle

\begin{abstract}
In atmospheric \v Cerenkov technique $\gamma-$rays are detected
against abundant background produced by hadronic showers. In order to
improve signal to noise ratio of the experiment, it is necessary to reject
a significant fraction of hadronic showers. The temporal and spectral
differences, the lateral distributions and density fluctuations of
\v Cerenkov photons generated by $\gamma-$ray and hadron primaries are
often used for this purpose. Here we study the differences in \v Cerenkov
photon density fluctuations at the observation level based on Monte Carlo
simulations. Various types of density fluctuations like the short range
(or local), medium range fluctuations and flatness parameter are studied. The
estimated quality factors reflect the efficiencies with which the hadrons
can be rejected from the data. It has been found that we can reject around
80\% of proton showers while retaining about 70\% of $\gamma-$ray showers
in the data, based only on the differences in the flatness parameter.
Density fluctuations
particularly suited for wavefront sampling observations seem to be a good
technique to improve the signal to noise ratio.
\end{abstract}

\section{Introduction}

In a typical wavefront sampling experiment, arrival time of \v Cerenkov
photons and \v Cerenkov photon density are sampled at several locations
in the \v Cerenkov pool generated by air showers initiated by $\gamma-$rays
from astronomical sources, using distributed array of telescopes.
Cosmic ray showers which also give rise to \v Cerenkov light similar
to that produced by $\gamma-$rays constitute abundant background against
which the $\gamma-$ray signal is to be detected. Hence it is necessary
to devise methods to reject a large fraction of cosmic ray background and
thereby improve the signal-to-noise ratio or sensitivity of the experiment.
Previously we have studied the usefulness of parameters based on timing
information recorded in wavefront sampling experiment for gamma-hadron
separation (Chitnis and Bhat, 2001). Use of arrival time jitter and
parameters based on shape of the \v Cerenkov pulse has been demonstrated.
Here we investigate the efficacy of certain parameters based on \v
Cerenkov photon density distribution for gamma-hadron separation.

\section{Local density fluctuations}

Pachmarhi Array of \v Cerenkov Telescopes (PACT) consists of 25
telescopes with each telescope consisting of para-axially mounted parabolic
mirrors of diameter 0.9 $m$ each (see Chitnis {\it et al.}, 2001 for details).
Here we study the usefulness of local density fluctuations or LDF for
gamma-hadron separation. LDF or density jitter is defined as the
ratio of RMS to mean of photon densities from 7 mirrors of each telescope.
We have simulated a large number of showers initiated by $\gamma-$rays of
energy 500 GeV and protons of energy 1 TeV, incident vertically at the
top of the atmosphere, for this purpose. Showers are simulated using CORSIKA
(Heck et al., 1998). An array of telescopes spread over an area of
400 $m$ $\times$ 400 $m$, much larger than PACT, is used for simulations. 
We have simulated 100 showers for each primary for each of the three
observation altitudes, $viz.$, sea level, 1 $km$ above sea level which 
corresponds to altitude of PACT and for 2.2 $km$ above sea level.

Figure 1 shows the variation of LDF as a function of distance from core of 
the shower for showers initiated by 500 GeV $\gamma-$rays and 1 TeV protons 
at various observation altitudes. It can be seen that the
LDF for protons is consistently higher than that for $\gamma-$ray
primaries for all altitudes and at all core distances. This is expected
due to differences in kinematics of these two types of showers (Chitnis and
Bhat, 1998). Also fluctuations are not very sensitive to core distances.
Hence LDF is a likely parameter to be used for gamma-hadron separation.

\begin{figure}[t]
\vspace*{2.0mm} 
\includegraphics[width=8.3cm]{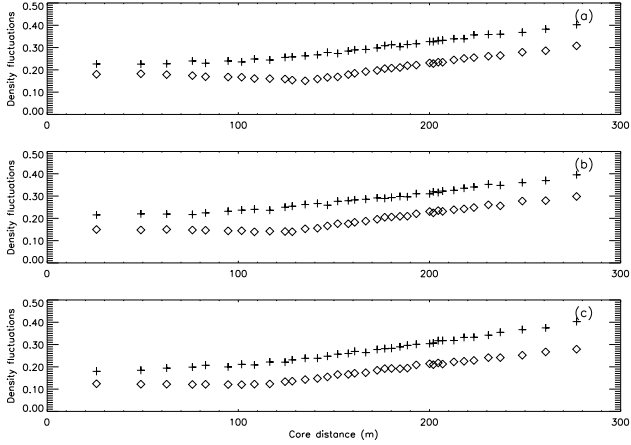}
\caption{Variation of local density fluctuations (LDF) as a function
of core distance averaged over 100 showers initiated by 500 GeV $\gamma-$rays
(indicated by diamond) and 1 TeV protons (indicated by +), for three different
altitudes: (a) sea level, (b) 1 km from sea level and (c) 2.2 km.}
\end{figure}

We use quality factor as a figure of merit of a parameter to distinguish
between $\gamma-$ray and proton initiated showers. It is defined as

$$ Q_f={{N_a^{\gamma}} \over {N_T^{\gamma}}} \left( {{N_a^{pr}} \over {N_T^{pr}}} \right) ^{-{{1} \over {2}}} \eqno $$

\noindent where $N_a^{\gamma}$ is the number of $\gamma-$rays accepted (i.e.
below threshold), $N_T^{\gamma}$ is the total number of $\gamma-$rays,
$N_a^{pr}$ is the number of protons accepted and $N_T^{pr}$ is the total
number of protons. Larger the quality factor, better is the background
rejection efficiency.

\begin{table}
\caption{Quality factors for local density fluctuations}
\begin{tabular}{lllll}
\hline
Obs & Threshold & Quality factor & Fraction  & Fraction \\
altitude  &value of &  & of  & of \\
  (km) & LDF & &  accepted & accepted \\
       &    &  & $\gamma-$rays & protons \\
\hline
0 & 0.248  & 1.284 $\pm$ 0.012  & 0.842 & 0.429 \\
1 & 0.257  & 1.215 $\pm$ 0.011  & 0.888 & 0.534 \\
2.2 & 0.234 & 1.261 $\pm$ 0.012 & 0.870 & 0.476 \\
\hline
\end{tabular}
\end{table}

Figure 2 shows the distribution of LDF from 500 GeV $\gamma-$rays and 1 TeV
protons, for three different altitudes. Optimum quality factors for each
of the cases are listed in Table 1. It can be seen that it is possible
to reject about 50\% proton showers retaining about 85\% of $\gamma-$ray
showers, using LDF.

\begin{figure}[t]
\vspace*{2.0mm} 
\includegraphics[width=8.3cm]{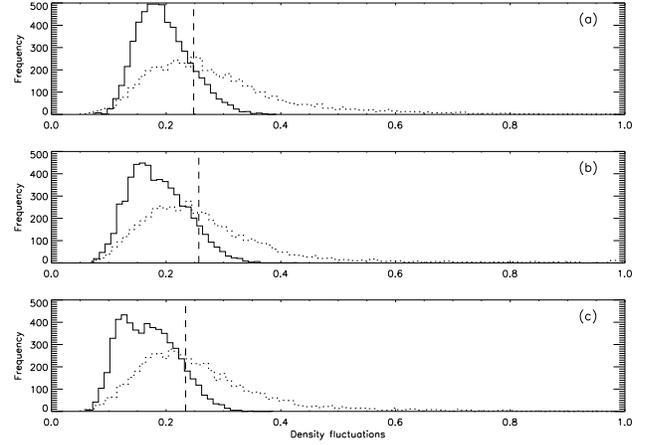}
\caption{Gamma-hadron separation based on LDF for 500 GeV $\gamma-$rays
(continuous line) and 1 TeV protons (dotted line) for observation altitudes of
(a) 0 km, (b) 1 km and (c) 2.2 km. Dashed line indicates the parameter value
which yields optimum quality factor (see table 1).}
\end{figure}

\section{Medium range density fluctuations}

PACT consists of four sectors of six telescopes each. We define medium
range or sector-wise density fluctuations (MDF) as the ratio of RMS to
mean density, where RMS and mean are calculated using total photon densities 
at each of the six telescopes of a sector. Figure 3 shows the variation of 
medium range density fluctuations as a function of core distance for showers 
initiated by 500 GeV $\gamma-$rays and 1 TeV protons for three different 
altitudes of observation level. As in the case of LDF using the distributions 
of MDF for showers initiated by 500 GeV $\gamma-$rays and 1 TeV protons, 
quality factors are calculated at three altitudes of observation levels. 
Optimum quality factors for different altitudes are listed in Table 2 and
distributions of MDF are shown in Figure 4. It can be seen that it is 
possible to reject about 60-70\% of proton showers retaining about 80\% of 
$\gamma-$ray showers, based on MDF.

\begin{figure}[t]
\vspace*{2.0mm} 
\includegraphics[width=8.3cm]{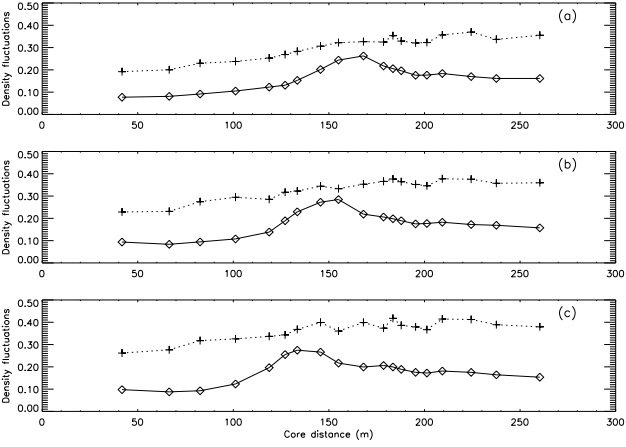}
\caption{Variation of medium range or sector-wise density fluctuations
as a function of core distance averaged over 100 showers initiated by 500 GeV
$\gamma-$ rays (indicated by diamond and continuous line) and 1 TeV protons
(indicated by plus sign and dotted line), for three different altitudes:
(a) 0 km, (b) 1 km and (c) 2.2 km}
\end{figure}

\begin{figure}[t]
\vspace*{2.0mm} 
\includegraphics[width=8.3cm]{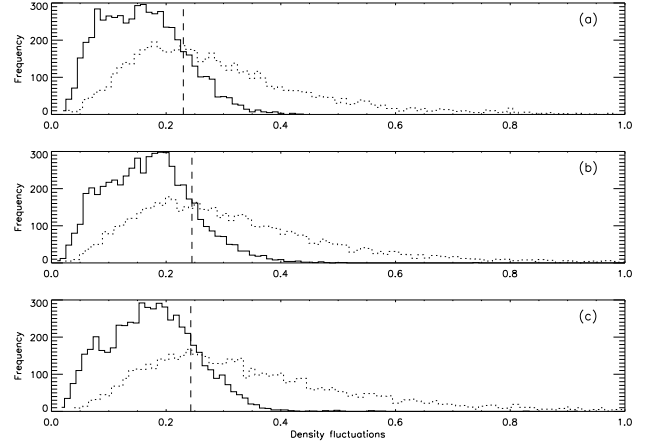}
\caption{Gamma-hadron separation based on medium range density
fluctuations for 500 GeV $\gamma-$rays (continuous line) and 1 TeV protons
(dotted line) for observation
altitudes of (a) 0 km, (b) 1 km and (c) 2.2 km. Dashed line indicates the
value of the parameter for optimum quality factor (see table 2).}
\end{figure}

\begin{table}
\caption{Quality factors for medium range density fluctuations}
\begin{tabular}{lllll}
\hline
Obs & Threshold & Quality factor & Fraction  & Fraction \\
altitude  & value of  &   & of  & of \\
  (km) & MDF & & accepted & accepted \\
  & & & $\gamma-$rays & protons \\   
\hline
0 & 0.230 & 1.297 $\pm$ 0.030 & 0.819 & 0.399 \\
1 & 0.245 & 1.332 $\pm$ 0.031 & 0.830 & 0.388 \\
2.2 & 0.243 & 1.418 $\pm$ 0.034 & 0.795 & 0.315 \\
\hline
\end{tabular}
\end{table}

\section{Medium range flatness parameter}

Lateral distributions of \v Cerenkov photons (variation of \v Cerenkov
photon density as a function of core distance) from showers initiated
by $\gamma-$rays show a characteristic hump at the core distance of about
120-140 m, depending on the observation altitude. This is due to the effective
focusing of \v Cerenkov photons from a large range altitudes. Distributions
are flat within the hump region and density falls rapidly beyond the hump.
Lateral distributions from proton showers, on the other hand, show
continuously falling density distribution as core distance increases (Rao 
and Sinha, 1988 and Chitnis and Bhat, 1998). Also due to the kinematical 
differences, the lateral distributions from $\gamma-$ray showers are smooth 
compared to proton showers. These differences in lateral distributions can 
be parameterized using flatness parameter defined as :

$$ \alpha={{1} \over {N}} \left[\sum_{i=1}^{N} {{ \left( \rho_i - \rho_0 \right)^2} \over {\rho_0}} \right] \eqno $$

\noindent where $N$ : no. of telescopes triggered, $\rho_i$ : photon density measured by
individual telescopes and $\rho_0$ : average density.
 
Lateral distributions from $\gamma-$ray showers are expected to have a
smaller value of $\alpha$ parameter compared to proton generated showers,
on the average. Figure 5 shows variation of $\alpha$ parameter
as a function of core distance for showers generated by 500 GeV
$\gamma-$rays and 1 TeV protons, for three observation altitudes.
It can be seen that, on an average,
$\gamma-$ray showers have smaller value of $\alpha$ compared to proton
showers. Also for both $\gamma-$rays and protons, value of $\alpha$
increases with increase in altitude of observation. At all the three
altitudes $\gamma-$ray showers show larger value of $\alpha$ near hump
region of lateral distribution. As a result, the difference in value of
$\alpha$ for $\gamma-$rays and protons near hump is reduced. Hence $\alpha$
can be useful discriminant at core distances away from the hump, on both
the sides.

\begin{table}
\caption{Quality factors for medium range flatness parameter for core
distance $<$ 100 m}
\begin{tabular}{lllll}
\hline
Obs & Threshold & Quality factor & Fraction  & Fraction \\
altitude   & value of  &   & of  & of \\
  (km) & $\alpha$  & & accepted  & accepted \\
       &  & & $\gamma-$rays & protons \\
\hline
0 & 0.33 & 1.800 $\pm$ 0.119 & 0.690 & 0.147 \\
1 & 0.52 & 1.556 $\pm$ 0.095 & 0.738 & 0.225 \\
2.2 & 0.66 & 1.662 $\pm$ 0.108 & 0.677 & 0.166 \\
\hline
\end{tabular}
\end{table}

\begin{figure}[t]
\vspace*{2.0mm} 
\includegraphics[width=8.3cm]{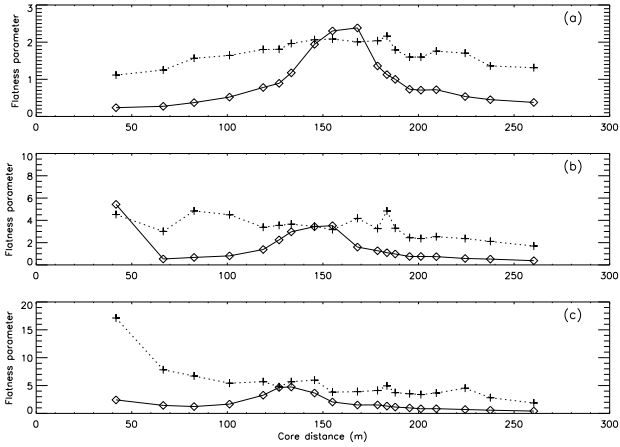}
\caption{Variation of $\alpha$ parameter as a function of core
distance for showers generated by 500 GeV $\gamma-$rays (diamonds and
continuous line) and 1 TeV protons (+ and dotted line), for altitude
of (a) 0 km, (b) 1 km and (c) 2.2 km. $\alpha$ parameter is calculated
for each sector.}
\end{figure}

Distributions of $\alpha$ parameter for telescopes within core distances of
100 $m$, for showers initiated by 500 GeV $\gamma-$rays and 1 TeV protons,
for observation altitude of 1 $km$ are shown in Figure 6. Optimum quality
factors for all the three altitudes are listed in Table 3. It can be seen
that the flatness parameter serves as a good discriminant for showers with
smaller impact parameters. For telescopes within 100 $m$ of shower axis
it is possible to reject about 80\% of the proton showers retaining
about 70\% of $\gamma-$ray showers based on flatness parameter alone.

\begin{figure}[t]
\vspace*{2.0mm} 
\includegraphics[width=8.3cm]{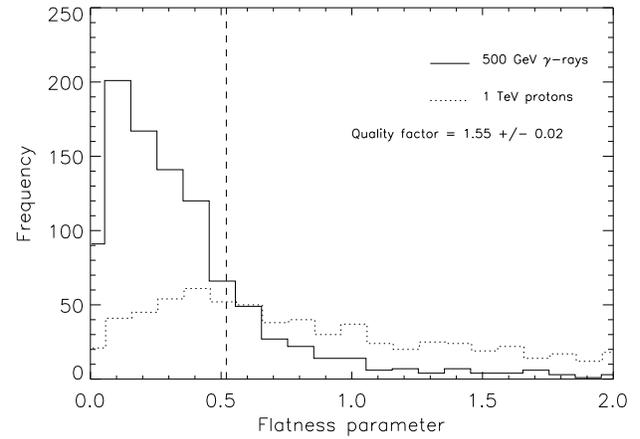}
\caption{Gamma-hadron separation based on $\alpha$ parameter
for sectors within 100 m from shower axis for showers initiated by
500 GeV $\gamma-$rays and 1 TeV protons, at observation altitude of 1 km.
Dashed line indicates the ordinate for optimum quality factor.
Quality factors for this case as well as for other observation altitudes
are listed in Table 3.}
\end{figure}

\section{Conclusions}

\begin{table}
\caption{Quality factors with density parameters applied in tandem, for core
distance $<$ 100 m}
\begin{tabular}{lllll}
\hline
Obs & Threshold & Quality factor & Fraction  & Fraction \\
altitude   & value of  &   & of  & of \\
  (km) & LDF, MDF  & & accepted  & accepted \\
       & \& $\alpha$ & & $\gamma-$rays & protons \\
\hline
0 & 0.23, 0.10 & 2.221 $\pm$ 0.167 & 0.655 & 0.087 \\
  & \& 0.33 &&&\\
1 & 0.19, 0.09 & 2.177 $\pm$ 0.175 & 0.580 & 0.071 \\
  & \& 0.52 &&&\\
2.2 & 0.15, 0.09 & 3.365 $\pm$ 0.368 & 0.563 & 0.028 \\
    & 0.66 &&&\\
\hline
\end{tabular}
\end{table}

In this work we have demonstrated the use of parameters based on \v
Cerenkov photon density fluctuations for gamma-hadron separation. Using
local density fluctuations it is possible to reject about 50\% of proton
showers retaining about 85\% of $\gamma-$ray initiated showers. Whereas,
based on medium range density fluctuations, it is possible to reject about
60-70\% of proton initiated showers retaining about 80\% of showers
produced by $\gamma-$rays. Flatness parameter, on the other hand, serves
as a useful discriminant at core distances away from hump. Using this
parameter it is possible to reject about 80\% of proton showers, retaining
about 70\% of $\gamma-$ray induced showers, for core distances within 100 $m$.
Using these three parameters in tandem it is possible to improve rejection
efficiencies further. Table 4 lists the quality factors at various altitudes 
obtained by applying all the three parameters, LDF, MF and $\alpha$ together.
Here we restrict to only the telescopes within the distance of 100 $m$ from
shower core. It can be seen that it is possible to reject about 95\%
of proton showers, retaining about 60\% of $\gamma-$ray showers at different
observation altitudes, using density based parameters in tandem. In addition,
using these parameters in tandem with the parameters based on timing
information such as timing jitter and \v Cerenkov pulse shape parameters
will greatly improve the sensitivity of the experiments based on wavefront
sampling technique.

\end{document}